\documentclass[prl,twocolumn,showpacs,amsmath,amssymb,superscriptaddress]{revtex4-1}
\usepackage{graphicx,amsfonts,amssymb,amsmath}
\usepackage{textcomp, color}
\usepackage[colorlinks,bookmarks=false,citecolor=blue,linkcolor=red,urlcolor=blue]{hyperref}

\begin{document}

\title{Critical Casimir forces from the equation of state of quantum critical systems}  
\author{Adam Ran\c{c}on}
\affiliation{Universit\'e de Lyon, ENS de Lyon, Universit\'e Claude Bernard, CNRS, 
Laboratoire de Physique, F-69342 Lyon, France} 
\author{Louis-Paul Henry}
\affiliation{Institute for Theoretical Physics, University of Innsbruck, A-6020 Innsbruck, Austria}
\author{F\'elix Rose} 
\affiliation{Laboratoire de Physique Th\'eorique de la Mati\`ere Condens\'ee, 
CNRS UMR 7600, Universit\'e Pierre et Marie Curie, 4 Place Jussieu, 
75252 Paris Cedex 05, France}
\author{David Lopes Cardozo}
\affiliation{Universit\'e de Lyon, ENS de Lyon, Universit\'e Claude Bernard, CNRS, 
Laboratoire de Physique, F-69342 Lyon, France} 
\author{Nicolas Dupuis} 
\affiliation{Laboratoire de Physique Th\'eorique de la Mati\`ere Condens\'ee, 
CNRS UMR 7600, Universit\'e Pierre et Marie Curie, 4 Place Jussieu, 
75252 Paris Cedex 05, France}
\author{Peter C. W. Holdsworth} 
\affiliation{Universit\'e de Lyon, ENS de Lyon, Universit\'e Claude Bernard, CNRS, 
Laboratoire de Physique, F-69342 Lyon, France} 
\author{Tommaso Roscilde}                    
\affiliation{Universit\'e de Lyon, ENS de Lyon, Universit\'e Claude Bernard, CNRS, 
Laboratoire de Physique, F-69342 Lyon, France} 
\affiliation{Institut Universitaire de France, 103 boulevard Saint-Michel, 75005 Paris, France}
\begin{abstract}
The mapping between a classical length and  inverse temperature as imaginary time  provides a direct equivalence between the Casimir force of a classical system in $D$ dimensions and internal energy of a quantum system in $d$$=$$D$$-$$1$ dimensions. The scaling functions of the critical Casimir force of the classical system with periodic boundaries thus  emerge from the analysis of the symmetry related quantum critical point.  We show that both non-perturbative renormalization group and quantum Monte Carlo analysis of quantum critical points provide quantitative estimates for the critical Casimir force in the  corresponding classical model, giving access to  widely different aspect ratios for the geometry of  confined systems. In the light of these results we propose protocols for the experimental realization of critical Casimir forces for periodic boundaries  through state-of-the-art cold-atom and solid-state experiments.
\end{abstract}

\def\calF{\mathcal{F}}
\def\calN{\mathcal{N}}
\def\Lp{L_\perp}
\def\inttau{\int_0^\beta d\tau}
\def\intr{\int d^dr}
\def\nablabf{\boldsymbol{\nabla}}
\def\phibf{\boldsymbol{\phi}}
\def\varphibf{\boldsymbol{\varphi}}
\def\llbrace{\left\lbrace}
\def\rrbrace{\right\rbrace}
\def\lbraket{\left[}
\def\rbraket{\right]}
\def\half{\frac12}
\def\dtau{{\partial_\tau}}  
\def\r{{\bf r}}
\maketitle

\emph{Introduction -} The confinement of a fluctuating field generates forces on the confining surfaces. This concept, originally developed by Casimir \cite{Casimir48} for the confinement of the quantum electromagnetic field between two perfectly conducting plates can be extended to classical fields and  critical phenomena in which diverging order parameter fluctuations at a second-order phase transition are cut off by the confinement \cite{deGennes78}. Being a critical phenomenon, this Casimir effect takes a form characterized by a universal scaling function that depends on both the bulk and the surface universality classes \cite{gambassi_casimir_2009,krech_casimir_1994}. Indeed, boundary conditions play a significant role in determining the scale and even the sign of the Casimir force as shown in both numerical \cite{Vasilyev2009,mohry_crossover_2010,hasenbusch_thermodynamic_2011,vasilyev_critical_2011}, and field theoretic studies. For the latter, effort has concentrated on the large $N$  expansions  (where $N$ is the number of components of the order parameter)  and the $\epsilon$-expansion close to four dimensions. The former allows analytic calculations over the whole phase diagram, but fails to catch the non-monotonous shape of the scaling function for periodic boundary conditions \cite{Danchev1996}. In the case of the $\epsilon$-expansion, it typically fails in the ordered phase, and converges  poorly at the critical point \cite{Diehl2006,Gruneberg2008}. 
On the experimental side, Casimir forces have so far been measured in wetting films of liquid Helium, driven through the lambda transition into the superfluid phase, or in binary mixtures showing a demixing transition in the dense fluid phase  \cite{fukuto_critical_2005,ganshin_critical_2006,garcia_critical_2002,hertlein_direct_2008}.

\begin{figure}[t]
\includegraphics[width=\columnwidth]{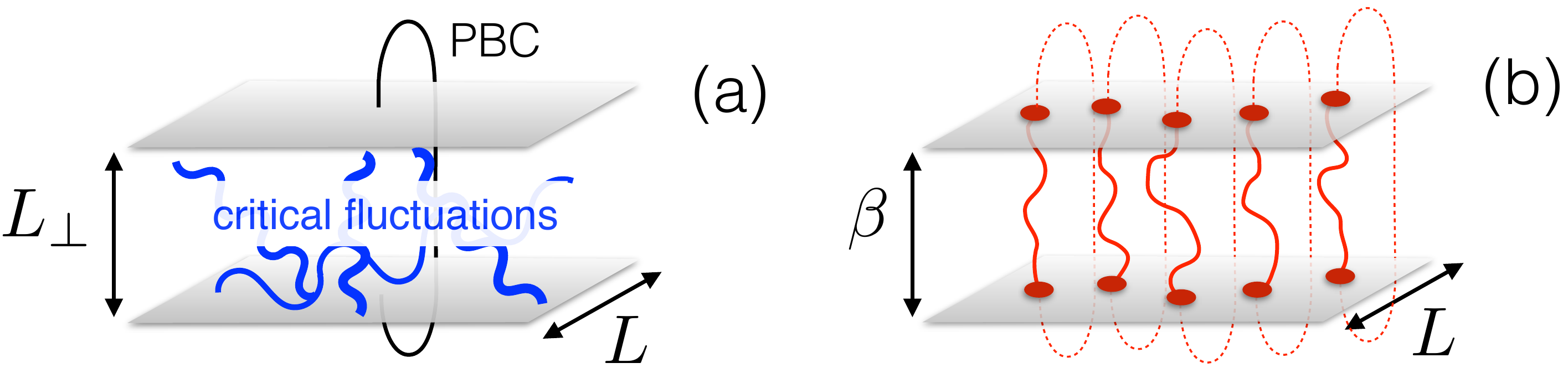}
\caption{The central theme of this paper is the correspondence between (a) $D$-dimensional classical critical phenomena in confined geometries with periodic boundary conditions (PBC); and (b) quantum many-body systems in $d=D-1$ dimensions at finite temperature and close to a quantum critical point, represented as effective classical $D$-dimensional systems.}
\label{fig_QCmapping}
\end{figure}
In the  functional integral description of quantum statistical physics, a $d$-dimensional  quantum system is represented by an effective classical system in $D$$=$$d$$+$$1$ dimensions with periodic boundary conditions in the extra dimension, the imaginary time, see Fig.~\ref{fig_QCmapping}.
This mapping has allowed for profound understanding of quantum many body systems in general and quantum critical points (QCP) in particular \cite{SachdevBook}. The general paradigm in this process has been for the classical mapping to provide insight and understanding of the quantum problem. 
 
 In this paper we turn this paradigm on its head, showing that mapping to the quantum system can provide new insight into the classical problem.
The critical Casimir force for a classical system with periodic boundary conditions appears naturally   in the  functional integral formulation close to a QCP \cite{DanchevBook}. We show that this generalized force emerges from the thermodynamics of the quantum system. As a consequence, we are able to 
turn the considerable arsenal of field theoretic and numerical tools developed for the QCP towards the critical Casimir scaling function. We show that the non-perturbative renormalization group (NPRG) provides  estimates for the scaling function of three-dimensional O($N$) spin models with unprecedented precision for a theoretical method. Further, we show that the scale of the critical Casimir force at the transition can be accurately estimated for the O($1$) model from quantum world-line Monte Carlo (QMC) simulations of the transverse field Ising model. This QMC analysis in two dimensions allows one to leave the slab geometry in three dimensions and explore a complete range of aspect ratios from thin film to columnar geometry \cite{Hucht2011}.

\emph{Classical versus quantum scaling functions -}
We consider a $D$-dimensional classical system with thickness $L_\perp$ and cross-sectional area $L^{D-1}$, which, in the thermodynamic limit, $L,L_\perp$$\to$$\infty$,  undergoes a second-order phase transition at a temperature $T_c$. The free energy can be written as 
\begin{equation}
\Omega(t,L,L_\perp) = L^{D-1} L_{\perp} k_B T [ \omega_{\rm ex}(t,L,L_\perp) + \omega_{\rm bulk}(t) ] , 
\label{Fcl} 
\end{equation}
where $t=$$(T$$-$$T_c)/T_c$ is the reduced temperature. Here $\omega_{\rm bulk}$ denotes the free energy density in the thermodynamic limit, in units of $k_BT$, and $\omega_{\rm ex}$ the ``excess'' contribution due to the finite volume of the system. For $D$$<$$4$, hyperscaling implies that the excess free energy density can be written in the scaling form~\footnote{It can be shown using renormalization-group arguments that $\omega_{\rm ex}$ has no regular part at the transition so that it
obeys the scaling form~(\ref{wexscaling}) (which would otherwise be satisfied only by its singular part).}
\begin{equation}
\omega_{\rm ex}(t,L,L_\perp) = L_\perp^{-D} \calF_\pm\left( \frac{L_\perp}{\xi},\frac{L_\perp}{L}\right) , 
\label{wexscaling}
\end{equation}
where $\calF_\pm$ is a universal scaling function which depends only on the universality class of the (bulk) phase transition and the boundary conditions. The $+$$/$$-$ index refers to the disordered ($T$$>$$T_c$) and ordered ($T$$<$$T_c$) phases, respectively. The correlation length $\xi$$=$$\xi_{0\pm}|t|^{-\nu}$ diverges at the transition with a critical exponent $\nu$. In the low-temperature phase, when the spontaneously broken symmetry is continuous, $\xi$ should be interpreted as the Josephson length, i.e. the length separating long-wavelength (gapless) Goldstone modes from critical fluctuations at shorter length scales~\cite{Josephson66}. The scaling form~(\ref{wexscaling}) holds whenever $\xi$, $L$ and $L_\perp$ are much larger than the Ginzburg length $\xi_G$ (scaling limit) \cite{SachdevBook}. 

\begin{table}[t!]
\caption{Conversion table between classical and quantum critical systems. $t$ is the reduced temperature of the classical system and $\delta$ the nonthermal control parameter of the quantum phase transition.}
\begin{center}
\begin{tabular}{cccccc}
\hline \hline
Classical & $ \quad D$ & $ \quad L_\perp$ & $ \quad t$ & $ \quad\xi$ & Casimir force \\ \hline
Quantum & $ \quad d+1$ & $ \quad\beta\hbar c$ & $  \quad\delta$ & $  \quad\xi,\, \xi_\tau $& \quad internal energy \\
\hline \hline
\end{tabular}
\end{center}
\label{tab_conv}
\end{table}

The Casimir force per unit area, in units of $k_BT$, is then \cite{dohm_critical_2009,Hucht2011}
\begin{equation}
\hspace{-0.3cm}f_C(t,L,L_\perp) =L_\perp^{-D} \vartheta\left( x,y\right) = - \frac{\partial}{\partial L_\perp} L_\perp \omega_{\rm ex}(t,L,L_\perp) ,
\label{fCdef} 
\end{equation}
where the choice of scaling variables, $x$$=$$t\left( L_\perp/\xi_{0+}\right)^{1/\nu}$ and $y$$=$$L_\perp/L  $ allows for the definition of a single universal scaling function above and below the transition ($\omega_{\rm ex}$$=$$L_\perp^{-D} \calF(x,y)$): 
\begin{equation}
\vartheta(x,y) = (D-1)\calF(x,y) - \frac{x}{\nu} \frac{\partial \calF(x,y)}{\partial x} - y \frac{\partial \calF(x,y)}{\partial y} . 
\label{Thetadef}
\end{equation}
The scale of the force is determined by $\vartheta(0,y)$, the value at the critical point. This amplitude passes through zero for $y$$=$$1$ and diverges as $y$$\to$$\infty$ \cite{Hucht2011}. It is thus useful, for $y$$>$$1$, to define $f_C$$=$$L^{-D}\tilde\vartheta$,  such that the scaling function $\tilde\vartheta(x,y)$$=$$y^{-D}\vartheta(x,y)$ has a finite limit for $x$$=$$0$ and $y$$\to$$\infty$.

\begin{figure*}[ht!]
\centerline{
\includegraphics[width=6cm]{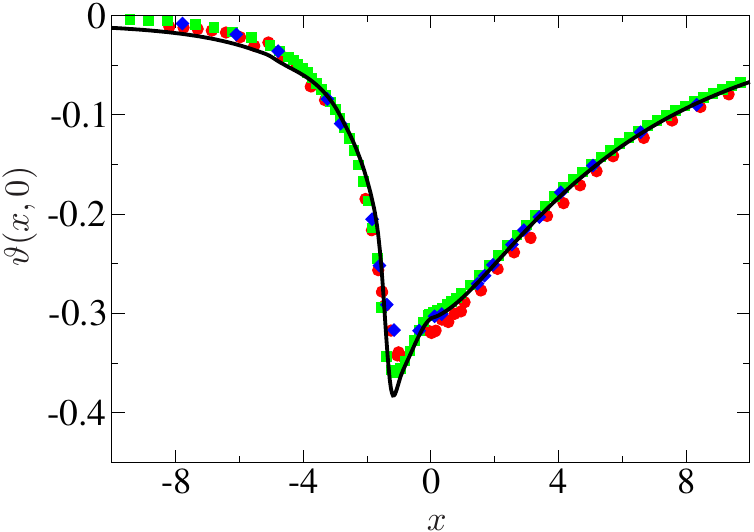}
\includegraphics[width=5.5cm]{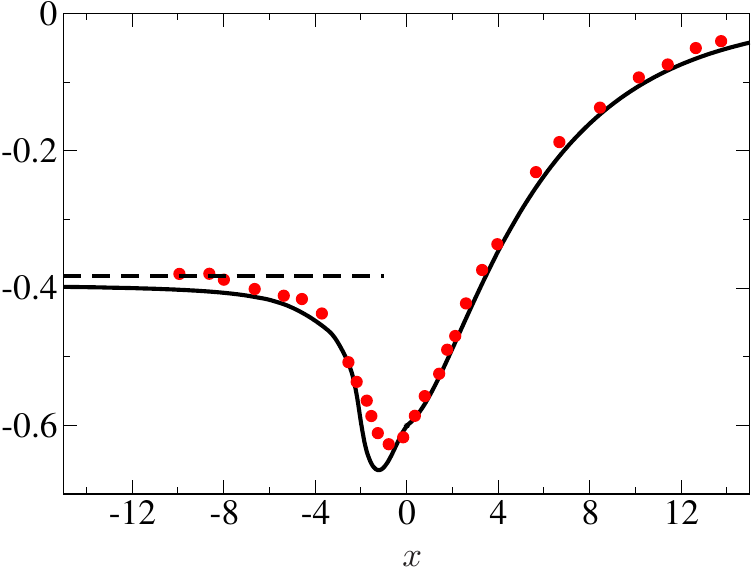}
\includegraphics[width=5.5cm]{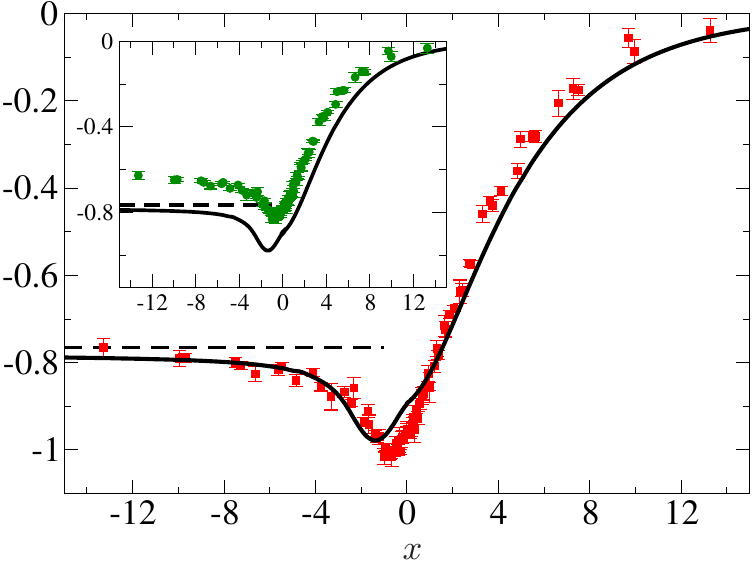}}
\caption{Casimir scaling function $\vartheta(x,0)$ for the three-dimensional O($N$) universality class from the NPRG approach to the two-dimensional quantum O($N$) model (full line), compared to classical Monte Carlo simulations of the corresponding three-dimensional spin models (symbols).  The horizontal dashed line shows the (exact) limit $-2(N$$-$$1$$)\zeta(3)/2\pi$.  (left panel)  Ising ($N$$=$$1$) universality class. Monte Carlo simulations are from Ref.~\cite{Vasilyev2009} (blue diamonds), Ref.~\cite{Hucht2011} (green squares) and Ref.~\cite{lopes_cardozo_finite_2015} (red circles). (middle panel)  XY universality class ($N$$=$$2$). The Monte Carlo data  are from Ref.~\cite{Vasilyev2009}. (right panel) Heisenberg universality class ($N$$=$$3$). The Monte Carlo  data~\cite{Dantchev2004} have been rescaled so as to satisfy the correct asymptotic value for $x$$\to$$-\infty$ (see text); the bare data are shown in the inset.}
\label{fig_N1N2N3}
\end{figure*}

The scaling function $\vartheta$ being universal, it is independent of the details of the microscopic interactions and is fully determined by the space dimension $D$, the nature of the order parameter, the range of the interactions and the boundary conditions. $\vartheta$ can therefore be computed from a field theory. For short-range interactions, the latter can be defined through the functional integral $Z=\int {\cal D}[\varphibf]\exp(-\beta H)$ for the partition function and the (local) Hamiltonian
\begin{equation} 
\beta H = \int_{L^{D-1}} d^{D-1}r_\parallel \int_0^{L_\perp} dr_\perp \,{\cal H}(\varphibf,\nablabf_\parallel\varphibf,\partial_\perp\varphibf;\lbrace g_i\rbrace) , 
\label{Hdef} 
\end{equation}
where the integration over ${\bf r}_\parallel$ is restricted to the area $L^{D-1}$. $\varphibf(\r_\parallel,r_\perp)$ denotes the $N$-component order parameter field and $\lbrace g_i\rbrace$ a set of coupling constants (we distinguish between parallel and perpendicular gradient terms, $\nablabf_\parallel\varphibf$ and $\partial_\perp\varphibf$, for later convenience). The dimensionless free energy, $\beta \Omega$$=$$-\ln Z$ where $\beta$$=$$1/k_BT$, depends on temperature through the (usually phenomenological) coupling constants of the classical field theory. For simplicity, we consider in the following homogeneous and isotropic classical systems.
 
To any such $D$-dimensional field theory defined in a volume $L^{D-1}L_\perp$ with periodic boundary conditions in the perpendicular direction, one can associate a quantum field theory in $d$$=$$D-1$ space dimensions by identifying $L_\perp$$\equiv$$\beta\hbar  c$, where $c$ is a characteristic velocity and where the spatial coordinate $r_\perp$$\equiv$$c\tau$ relates to an imaginary time $\tau$. 
The Hamiltonian of the classical theory maps onto the (Euclidean) action of the quantum field theory,
\begin{equation}
S = \int_0^{\hbar\beta} d\tau \int_{L^d} d^dr \, {\cal H}(\varphibf,\nablabf\varphibf,\dtau\varphibf;\lbrace g_i\rbrace) , 
\label{Squ}
\end{equation}
where the $g_i$'s are now temperature independent. Although $\beta H$ and $S$ are formally identical (up to a change in notations) they describe different physical systems. The Hamiltonian $H$ describes a $D$-dimensional classical system which undergoes a thermal phase transition in the thermodynamic limit $L,L_\perp$$\to$$\infty$. The action $S$ describes a $d$-dimensional quantum system which, in the thermodynamic limit $L$$\to$$\infty$, undergoes a zero-temperature phase transition where both the correlation length $\xi$$=$$\tilde \xi_{0,\pm} |\delta|^{-\nu}$ and the time scale $\xi_\tau$$=$$\xi/c$ diverge, and the critical modes at the QCP have a linear dispersion, $\omega$$=$$c|{\bf q}|$, corresponding to dynamical exponent $z$$=$$1$ \cite{SachdevBook}. This transition is driven by a non-thermal parameter $\delta$, assumed here to vanish at the QCP, which enters in $S$ only through the (usually phenomenological) $\delta$ dependence of the coupling constants $g_i$.

The critical point described by the classical field theory and the QCP described by the quantum field theory are formally equivalent and fall in the same universality class. A finite area $L^{D-1}$ in the classical model corresponds to a finite volume $L^d$ in the quantum model, and a finite thickness $L_\perp$ to a nonzero temperature $T$. The scaling analysis of the classical model straightforwardly translates to the quantum model. From Eqs.~(\ref{Fcl}, \ref{wexscaling}), we obtain the free energy
\begin{align}
\Omega(\delta,L,T) ={}& \Omega_{\rm bulk} + L^d \frac{(k_BT)^{d+1}}{(\hbar c)^d} \calF\left(x,y\right) ,
\label{Fqu}
\end{align} 
where $\Omega_{\rm bulk}$$=$$L^d\epsilon_{\rm gs}(\delta)$ is the zero-temperature bulk contribution, proportional to the ground-state energy density $\epsilon_{\rm gs}$. The scaling variables are now $x$$=$$\delta (\beta \hbar c/\tilde \xi_{0,+})^{1/\nu}$ and $y$$=$$\beta\hbar c/L$.
The internal energy density $\epsilon$$=$$L^{-d}\partial(\beta \Omega)/\partial\beta$ is given by  
\begin{equation} 
\begin{split}
\epsilon(\delta,L,T) ={}& \epsilon_{\rm gs}(\delta) - \frac{(k_BT)^{d+1}}{(\hbar c)^d} \vartheta \left(x,y \right) ,
\label{eps}
\end{split}
\end{equation}
where  $\vartheta$ is the universal scaling function of the critical Casimir force defined in Eq.~(\ref{Thetadef}). 
Comparing Eqs.~(\ref{fCdef}) and (\ref{eps}), we see that the Casimir force provides a measure of the difference between $\epsilon_{\rm gs}$ and $\epsilon(\delta,L,T)$. Notably, taking the thermodynamic limit, $L$$\to$$\infty$, $(y$$=$$0)$, one can deduce from this analysis, without prior knowledge, that the critical Casimir force of the classical system in slab geometry is negative, given that in this limit $-\vartheta$ is proportional to the thermal energy, which is always positive. The situation for general $y$ is discussed further below.
A summary of the conversion from the classical to the quantum terminology is given in Table~\ref{tab_conv}.

\emph{Renormalization group calculation of the critical Casimir force in O($N$) models -} 
The two-dimensional quantum O($N$) model is defined by the action 
\begin{align}
\hspace{-0.3cm}S ={}& \int_0^{\hbar\beta} d\tau \int d^2 r \biggl\lbrace \frac{(\nablabf\varphibf)^2}{2} + \frac{(\dtau \varphibf)^2}{2c^2}  + \frac{r \varphibf^2}{2} + \frac{u{(\varphibf^2)}^2}{4!} \biggr\rbrace ,  
\label{action1} 
\end{align}
where $\varphibf$ is an $N$-component real field satisfying periodic boundary conditions $\varphibf(\r,\tau+\hbar\beta)$$=$$\varphibf(\r,\tau)$. $r$ and $u$ are temperature-independent coupling constants and $c$ is the (bare) velocity of the $\varphibf$ field. 
The QCP at $r$$=$$r_{c}$ ($\delta$$=$$r$$-$$r_{c}$ for this model)  is in the universality class of the three-dimensional classical O($N$) model, and the phase transition is governed by the three-dimensional Wilson-Fisher fixed point. 

The renormalization group is a natural approach to compute universal quantities in the (quantum) O($N$) model. The calculation of scaling functions of the (2+1)-dimensional Wilson-Fisher fixed point is however notoriously difficult and perturbative renormalization group usually fails. In the following, we show that the  NPRG  provides us with a scaling function of the critical Casimir force which compares very well with results obtained from Monte Carlo simulations of three-dimensional classical spin systems (see also~\cite{Jakubczyk2013}). We only consider the thermodynamic limit, i.e. $L$$\to$$\infty$, and thus the scaling function $\vartheta(x,0)$. 

The NPRG is an implementation of the Wilsonian RG based on an exact equation for the Gibbs free energy (or ``effective action" in the field theory terminology) for which powerful approximation schemes have been designed~\cite{Berges2002,Delamotte2012}. Recently, the NPRG has been used to study the thermodynamics of the quantum O($N$) model~\cite{Rancon2013}, the Higgs amplitude mode~\cite{Rancon2014,Rose2015} and the quantum-to-classical crossover in the dynamics~\cite{Mesterhazy2015}. 
Our results, which are exact in the limit $N$$\to$$\infty$, are obtained from a derivative expansion of the effective action to second order and improves on the approach of Ref.~\cite{Rancon2013} (see the Supplemental Material for more details).

Figure~\ref{fig_N1N2N3}   shows the Casimir scaling function $\vartheta$ obtained from the two-dimensional quantum O($N$) model within the NPRG approach for the three-dimensional Ising ($N$$=$$1$), XY ($N$$=$$2$) and Heisenberg ($N$$=$$3$) universality classes, 
together with data from Monte Carlo simulations of the three-dimensional classical spin systems~\cite{lopes_cardozo_finite_2015,Hucht2011,Vasilyev2009,Dantchev2004}. In all cases we find very good agreement between the NPRG and simulation results. In particular, the non-monotonous form for $\vartheta(x,0)$ is well reproduced and the amplitude and position of the minimum of the scaling function are accurately predicted, with some small differences between NPRG and simulations occurring in the region around $x$$\simeq$$-1$, with the former showing a more pronounced minimum  for $N$$=$$1$ and $N$$=$$2$. 
Note that in Ref.~\cite{Dantchev2004}, the overall scale of the $N$$=$$3$ scaling function was not determined. We have rescaled the MC data so that they satisfy the known asymptotic value when $x$$\to$$-\infty$$, -2(N$$-$$1$$)\zeta(3)/2\pi$, corresponding to the excess free energy of bosons with linear dispersion~\cite{Rancon2013}; the rescaled function compares well with the NPRG result. 

We show in Table~\ref{tab_Pic} the NPRG and Monte Carlo estimates for the universal Casimir amplitude $\vartheta(0,0)/2$. Again the NPRG results are in very good agreement with MC simulations, with a relative difference below 1\% \footnote{No estimate was given in Ref.~\cite{Dantchev2004} ($N$$=$$3$) since the absolute amplitude of the Casimir force was not determined. }.

\begin{table}
\caption{Universal Casimir amplitude $\vartheta(0,0)/2$.}
\begin{center}
\begin{tabular}{cccc}
\hline \hline
$N$ & 1 & 2 & 3 \\ \hline
NPRG& $-0.1527$ & $-0.3006$ & $-0.4472$ \\
Monte Carlo~\cite{Vasilyev2009} & \quad $-0.1520(2)$ & $ \quad-0.2993(7)$ \\
\hline \hline
\end{tabular}
\end{center}
\label{tab_Pic}
\end{table}

\emph{Finite-size scaling and aspect ratio -}
The method of choice for a fully quantitative study of a QCP is  QMC. It provides the flexibility to vary the spatial as well as the time dimension, allowing for the evolution from slab to column geometry in the corresponding classical system through the variation of the ratio $y$$=$$\beta\hbar c/L$. As a consequence, the standard finite-size and finite-temperature scaling analysis of the numerical results close to the quantum critical point can be re-cast in the language of critical Casimir forces in columnar geometry \cite{DanchevBook}. Indeed the finite nature of the simulation cell implies that the quantum limit, $\beta$$\to$$\infty$, corresponds to column geometry for the corresponding classical system. 
With continuous-time QMC \cite{Prokofev}, the imaginary time axis becomes a continuous periodic dimension of length $\beta \hbar c$, as in the field theoretic approach, so that $y$ can be easily tuned to any value. Furthermore, the internal energy $\epsilon$$=$$L^{-d}\langle \hat H \rangle$, where $ \hat H$ is the Hamiltonian of the quantum system, is an easily accessible  observable, whereas the numerical methods for calculating the Casimir force in classical systems are computationally intensive \cite{Vasilyev2009,Hucht2011,hasenbusch_thermodynamic_2011,LopesCardozo2014}.
Following Eq.~(\ref{eps}) one can fit  the numerical calculation for energy density $\epsilon(\delta,L,T)$ to a suitable scaling function (this kind of fit has been used to compute $\tilde\vartheta$ in the limit $y\gg1$ for quantum systems  \cite{Sandvik1997,Hamer2000}).

\begin{figure}[t!]
\includegraphics[width= 6cm]{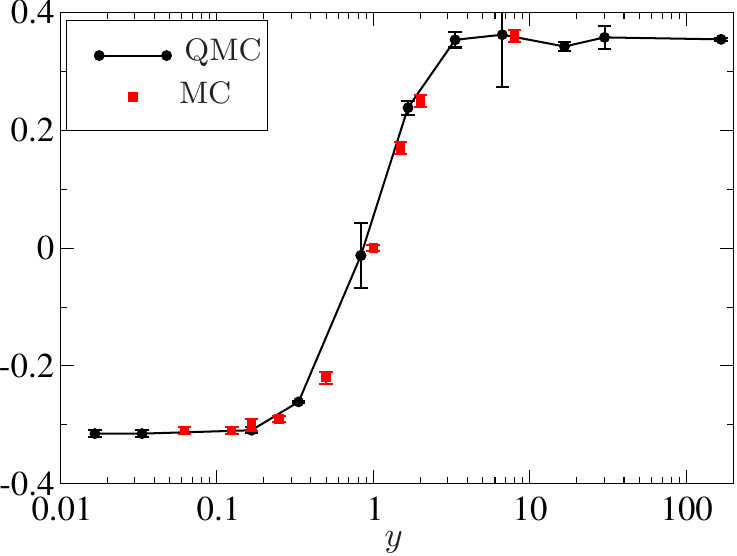}
\caption{Evolution of the Casimir amplitude extracted  using QMC from the quantum Ising model as a function of the aspect ratio $y$. We show $\vartheta(0,y)/2$ for $y\leq 1$ and $\tilde \vartheta(0,y)$ for $y\geq 1$. (Red) squares are classical Monte Carlo (MC) data from Ref.~\cite{Hucht2011}.}
\label{fig_aspect}
\end{figure}

We have studied the quantum critical point of the transverse-field Ising model in two-dimensions,
$	\hat H$$=$$-J\sum_{\langle i,j\rangle}\hat \sigma_i^z\hat \sigma_j^z$$-$$h\sum_i\hat \sigma_i^x ,$
which has a QCP at $h$$=$$h_c$ ($\delta$$=$$h-h_c$ here) between a ferromagnetic and a paramagnetic ground-state.
We have used a cluster QMC algorithm \cite{Blote2002} to compute the energy density, while the critical velocity $c$ is extracted from the excitation spectrum  at the QCP \cite{Schuler2016}, see Supplemental Material.
Numerical results for the critical Casimir amplitude estimated from QMC over the range $0$$<$$y$$<$$\infty$ are shown in Fig. \ref{fig_aspect} and compared with classical simulation results for the three-dimensional Ising model from Ref.~\cite{Hucht2011}. Excellent agreement is found, confirming the equivalence of these two critical phenomena away from the limit of slab geometry. 
There is a  sign change at $y$$=$$1$: for $y$$\ll$$1$, $-\vartheta$ probes the (positive-definite) thermal energy density, whereas for $y$$\gg$$1$ it probes the finite-size corrections to the ground-state energy density, which are usually negative for quantum systems.

\emph{Conclusion -}
The finite-temperature equation of state for a quantum critical system in dimension $d$ can in principle be measured in state-of-the-art experiments on quantum critical phenomena, including trapped ions \cite{Monroeetal2015} and quantum Ising magnets in a transverse field  \cite{Coldeaetal2010} for $N$$=$$1$; ultracold Bose gases loaded in optical lattices for $N$$=$$2$ \cite{Blochetal2008}; and quantum magnets under pressure for $N$$=$$2$ and $N$$=$$3$ \cite{Rueggetal2008}.  The critical Casimir force for a classical system in dimension $D$ with a thermal critical point could hence be experimentally accessed, opening the door to a new class of critical Casimir force experiments in which the quantum system becomes a simulator for confinement effects on critical fluctuations at a classical critical point. This approach would naturally put periodic boundaries on the experimental map for the first time, providing new motivation for detailed theoretical and numerical analysis of such systems.  The measure of the thermal energy can be achieved in the solid-state context via temperature integration of the specific heat; and in the atomic physics context by direct measurement of spin-spin or density-density correlation functions for the potential part, and by time-of-flight measurement for the kinetic part. Even though these measurements may reach the limits of the present experimental technologies, we hope that our work will provide a catalyst for future developments in this exciting direction.

\emph{Acknowledgements - } We thank A. Gambassi, A. Hucht and D. Dantchev for sharing the data of Refs. \cite{Vasilyev2009,Hucht2011,Dantchev2004}. It is a pleasure to thank F. Alet and S. Ciliberto for useful discussions.  DLC and PCWH acknowledge financial support from the ERC grant OUTEFLUCOP and the numerical resources of the PSMN at the ENS Lyon. This work is supported by ANR (``ArtiQ" project).

%



\def\calF{\mathcal{F}}
\def\calN{\mathcal{N}}
\def\Lp{L_\perp}
\def\inttau{\int_0^\beta d\tau}
\def\intr{\int d^dr}
\def\nablabf{\boldsymbol{\nabla}}
\def\phibf{\boldsymbol{\phi}}
\def\varphibf{\boldsymbol{\varphi}}
\def\llbrace{\left\lbrace}
\def\rrbrace{\right\rbrace}
\def\lbraket{\left[}
\def\rbraket{\right]}
\def\half{\frac12}
\def\dtau{{\partial_\tau}}  
\def\r{{\bf r}}
\def\w{\omega} 
\def\wn{\omega_n} 
\def\dk{\partial_k} 
\newcommand{\Tr}{{\rm Tr}}

\setcounter{equation}{0}
\newpage
\section{Supplemental Material \\ 
Critical Casimir forces from the equation of state of quantum critical systems}

\subsection{I. NPRG approach to the quantum O($N$) model}

The strategy of the NPRG approach is to build a family of models indexed by a momentum scale $k$ such that fluctuations are smoothly taken into account as $k$ is lowered from a microscopic scale $\Lambda$ down to 0~\cite{Berges2002a,Delamotte2012a}. In the case of the $d$-dimensional quantum O($N$) model, this is achieved by adding to the action the infrared regulator term 
\begin{equation}
\Delta S_k[\varphibf] = \half \sum_q R_k(q) \varphibf(-q) \cdot \varphibf(q) ,    
\end{equation}
where $q=({\bf q} ,i\wn)$ with ${\bf q} $ a $d$-dimensional momentum and $\wn=2n\pi T$ ($n$ integer) a bosonic Matsubara frequency. The partition function 
\begin{equation}
{\cal Z}_k[{\bf J} ] = \int {\cal D}[\varphibf] \, e^{-S[\varphibf] - \Delta S_k[\varphibf] + \inttau \intr {\bf J} \cdot\varphibf } 
\end{equation}
is now $k$ dependent. The scale-dependent effective action (or Gibbs free energy) 
\begin{equation}
\Gamma_k[\phibf] = - \ln {\cal Z}_k[{\bf J}] + \inttau \intr\, {\bf J} \cdot\phibf  - \Delta S_k[\phibf] 
\end{equation}
is defined as a (slightly modified) Legendre transform of the free energy $-\ln Z_k[{\bf J}]$ which includes the subtraction of $\Delta S_k[\phibf]$. Here $\phibf(\r,\tau)=\langle\varphibf(\r,\tau)\rangle$ is the order parameter (in the presence of the external source ${\bf J} $). Assuming that fluctuations are completely frozen by the $\Delta S_k$ term when $k=\Lambda$, mean-field theory becomes exact and $\Gamma_\Lambda[\varphibf]=S[\varphibf]$. On the other hand, the effective action of the original model is given by $\Gamma_{k=0}$ provided that $R_{k=0}$ vanishes. For a generic value of $k$, the cutoff function $R_k$ suppresses fluctuations with momentum $|{\bf q} |\leq k$ or frequency $|\wn|\leq c k$ but leaves unaffected those with $|{\bf q} |,|\wn|/c\geq k$ ($c$ denotes the velocity of the $\varphibf$ field). The variation of the effective action with $k$ is given by Wetterich's equation~\cite{Wetterich1993a} 
\begin{equation} 
\dk \Gamma_k[\phibf] = \half \Tr \Bigl\lbrace \dk R_k \bigl( \Gamma^{(2)}_k[\phibf] + R_k \bigr)^{-1} \Bigr\rbrace ,
\label{eqwet} 
\end{equation}
where $\Gamma_k^{(2)}$ denotes the second-order functional derivative of $\Gamma_k$. In Fourier space, the trace involves a trace over momenta and frequencies as well as the O($N$) index of the $\phibf$ field. 

To solve Eq.~(\ref{eqwet}), we use a derivative expansion of the scale-dependent effective action with the usual (Lorentz-invariant) exponential cutoff function~\cite{Berges2002a}. Such an expansion is made possible by the regulator term $\Delta S_k$ which ensures that all vertices $\Gamma^{(n)}_k$ are smooth functions of momenta ${\bf q}_i$ and frequencies $\omega_{n_i}$ and can be expanded in powers of ${\bf q}^2_i/k^2$ and $\w^2_{n_i}/(ck)^2$ when $|{\bf q}_i|,|\w_{n_i}|/c\ll k$. The derivative expansion of the effective action is fully determined by the O($N$) symmetry of the model. To second order, 
\begin{align}
\Gamma_k[\phibf] ={}& \int_0^{\hbar\beta} d\tau \intr \biggl\lbrace \frac{Z^x_k(\rho)}{2} (\nablabf\phibf)^2 +  \frac{Z^\tau_k(\rho)}{2} (\dtau\phibf)^2 \nonumber \\ & 
+ \frac{Y^x_k(\rho)}{4} (\nablabf\rho)^2 + \frac{Y^\tau_k(\rho)}{4} (\dtau\rho)^2 + U_k(\rho) \biggr\rbrace , 
\label{gamde} 
\end{align}
where we have introduced the O($N$) invariant $\rho=\phibf^2/2$. At zero temperature, Lorentz invariance of the quantum O($N$) model implies that $Z^x_k=Z^\tau_k$ and $Y^x_k=Y^\tau_k$ (this is not true anymore at nonzero temperatures). 

All thermodynamic quantities can be obtained from the effective potential
\begin{equation}
U_k(\rho) = \frac{1}{\beta L^d} \Gamma_k[\phibf] \Bigl|_{\phibf={\rm const}}  
\end{equation}
obtained from the effective action computed in a constant, i.e. uniform and time-independent, field. In particular, the free energy of the system is simply 
\begin{equation}
\Omega = L^d \lim_{k\to 0} U_k(\rho_{0,k}) , 
\label{Fdef} 
\end{equation}
where $\rho_{0,k}$ denotes the position of the minimum of the effective potential ($\rho_{0,k}$ vanishes in the disordered phase while $|\langle\varphibf\rangle|=\sqrt{2\rho_{0,k}}$ in the ordered phase). Inserting the ansatz~(\ref{gamde}) into (\ref{eqwet}) we obtain coupled equations for the $\rho$-dependent functions $U_k$, $Z^x_k$, $Z^\tau_k$, $Y^x_k$ and $Y^\tau_k$, which can be solved numerically. One can thus obtain the free energy $\Omega(T,r)$ using~(\ref{Fdef}) and in turn the universal scaling functions $\calF$ and $\vartheta$. 

The scaling function $\calF$ has been computed in Ref.~\cite{Rancon2013a} using a simpler approximation where the functions $Z^x_k(\rho)$ and $Z^\tau_k(\rho)$ are set equal to $Z^x_k(\rho_{0,k})$ and $Z^\tau_k(\rho_{0,k})$, the functions $Y^x_k(\rho)$ and $Y^\tau_k(\rho)$ omitted, and the effective potential $U_k(\rho)$ expanded to quadratic order. The complete derivative expansion [Eq.~(\ref{gamde})] that we use to compute $\calF$ and $\vartheta$ is exact is the limit $N\to\infty$. Furthermore, this expansion is known to be very efficient for small values of $N$. For the classical O($N$) model, it recovers not only the one-loop result near four dimensions but also the one-loop results near two dimensions obtained from the nonlinear sigma model~\cite{Berges2002a,Delamotte2012a}, and provides accurate estimates of the critical exponents in three dimensions~\cite{Balog16}. 
It also recovers quantitatively the main universal features of the Kosterlitz-Thouless transition in the two-dimensional O(2) model~\cite{Gersdorff2001,Jakubczyk2014,Jakubczyk2016}. In the two-dimensional quantum O($N$) model recent calculations~\cite{Rose2016} of the $T=0$ universal ratio $\Delta/\rho_s$ between the excitation gap in the disordered phase and the stiffness in the ordered phase are in very good agreement with Monte Carlo simulations~\cite{Gazit2013} for $N=2$ and $N=3$. 

\subsection{Quantum Monte-Carlo simulations}
We compare the results of the Casimir force of the classical 3D Ising model to that of the transverse-field Ising model 
\begin{equation}
\hat H=-J\sum_{\langle i,j\rangle}\hat \sigma_i^z\hat \sigma_j^z-h\sum_i\hat \sigma_i^x,
\end{equation}
which is critical for $h=h_c=3.04438$ \cite{Blote2002a}. This model is simulated via  a continuous-time world-line Monte-Carlo scheme, supplemented with a cluster update  \cite{Blote2002a} to overcome critical slowing down at the quantum phase transition.

  For our computations, we used system sizes ranging from $L=16$ up to $L=72$ (for the smallest values of the aspect ratio). For each system, the average energy was computed from several (from 16 to 256) independent runs of $10^4$ to $10^6$ measurements. Between two measurement, the number of cluster updates $n_c$  was chosen so that $n_c\langle s_c\rangle\gtrsim \beta L^2$, where $\langle s_c\rangle$ is the average cluster size. This leads to an autocorrelation-time of order one Monte Carlo step or less.
    The world-line Monte-Carlo allows to extract the excitation spectrum of the system, through the evaluation of the imaginary-time spin-spin correlation function. 
    Indeed, the spin-spin correlation function at momentum $\mathbf{q}$ is given by
    \begin{equation}
      \begin{array}{ll}
	S^{zz}(\mathbf{q},\tau)&=\dfrac{1}{\beta N}\left\langle\displaystyle\sum_{i,j}{\displaystyle\int_0^{\beta}{d\tau'\,e^{-\mathrm{i}{\bf q}\cdot({\bf r}_i-{\bf r}_j)}s_i^z(\tau')s_j^z(\tau'+\tau)}}\right\rangle\\
	&\underset{\beta\to\infty,\tau\to\infty}{\approx} e^{-\Delta_\mathbf{q} \tau},
      \end{array}
    \end{equation}
    where $\Delta_\mathbf{q}$ is the excitation gap at momentum $\mathbf{q}$. This allows to extract the lower band of the excitation which, in the limit $|\mathbf{q}|$$\to$$0$, is given by 
    \begin{equation}
      \Delta_\mathbf{q}=c |\mathbf{q}| ,
    \end{equation}
    with $c$ the critical velocity. One can then extract the velocity from the spectrum, see Ref. \cite{Schuler2016a} for a more detailed discussion.

\end{document}